\documentstyle [12pt]{article}

\begin{document}

\vspace{2mm}

\begin{flushright}
Preprint MRI-PHY/7/95 \\

hep-th/9506020, June 1995 
\end{flushright}

\vspace{2ex}

\begin{center}
{\large \bf Some Cosmological Consequences of Non \\

\vspace{2ex} 

            Trivial PPN Parameters $\beta$ and $\gamma$ } \\

\vspace{6mm}
{\large  S. Kalyana Rama}
\vspace{3mm}

Mehta Research Institute, 10 Kasturba Gandhi Marg, 

Allahabad 211 002, India. 

\vspace{1ex}
email: krama@mri.ernet.in \\ 
\end{center}

\vspace{4mm}

\begin{quote}
ABSTRACT. We study homogeneous isotropic universe in a \\
graviton-dilaton theory obtained, in a previous paper, by a simple 
requirement that the 
theory be able to predict non trivial values for $\beta$ and/or 
$\gamma$  for a charge neutral point star, without any naked 
singularities. We find that in 
this universe the physical time can be continued indefinitely into 
the past or future, and that all the physical curvature invariants are 
always finite, showing the absence of big bang singularity. Adding 
a dilaton potential, we find again the same features. As a surprising 
bonus, there emerges naturally a Brans-Dicke function, which has 
precisely the kind of behaviour needed to make 
$\omega_{bd} ({\rm today}) > 500$ in hyperextended inflation. 

\end{quote}

\newpage

\vspace{4ex}

{\bf 1.} For various reasons, as given in \cite{will}, it is worthwhile 
to consider alternative theories of gravity other than Einstein's. Two of 
the popular ones are Brans-Dicke (BD) theory \cite{dicke} and low energy 
string theory, which contain an extra scalar field $\phi$, called 
dilaton. Such alternative theories can be distinguished by the 
parametrised post Newtonian (PPN) parameters. Two of them, $\beta$ and 
$\gamma$ (both equal to one in Einstein's theory), can be 
calculated from static spherically symmetric solutions. 

By studying such solutions, it is shown in \cite{kppn} that neither BD 
nor low energy string theory can predict non trivial values 
of the PPN parameters $\beta$ 
and/or $\gamma$ for a charge neutral star, without any naked
singularities. In \cite{knobh} we looked for, and obtained, 
a graviton-dilaton theory with an arbitrary function $\psi(\phi)$, 
requiring it only to be able to 
predict non trivial values for $\beta$ and/or $\gamma$  for 
a charge neutral point star, without any naked singularities. These 
requirements imposed certain constraints on $\psi$, described 
below. For the static spherically symmetric case, these constraints led 
to the novel features of the gravitational force becoming repulsive at 
distances of the order of, but greater than, the Schwarzschild radius 
$r_0$ and the absence of a horizon for $r > r_0$. These results suggest 
that black holes are unlikely to form in a stellar collapse in this 
theory. See \cite{knobh} for further details. 

In this letter, we study solutions describing a homogeneous isotropic 
universe in the graviton-dilaton theory, obtained as above. 
In the case of BD or 
low energy string theory, the homogeneous isotropic universe has a big 
bang singularity. By this, we mean here that there is a curvature 
singularity at a finite time in the past, beyond which 
the physical time cannot be continued. However, in the case of the 
graviton-dilaton theory considered here, the constraints on $\psi$  
beautifully ensure that the physical time $\hat{t}$ can be continued 
indefinitely into the past or future, and that 
all the physical curvature invariants are finite for all $\hat{t}$. 
This shows that the big bang singularity is absent here. 

We next add to the action a dilaton potential $V(\phi)$, 
made finite for all $\phi$ by restricting the range of a parameter. 
For the homogeneous isotropic universe in this theory, with 
$V$ present, we again find that the physical time $\hat{t}$ can be 
continued indefinitely into the past or future, and that 
all the physical curvature invariants are finite for all $\hat{t}$. 
This shows that the big bang singularity is absent here too.  

The theory considered here has another surprising feature which 
emerges naturally. As described below, the graviton-dilaton action can 
be rewritten, containing now a BD function 
$\omega_{bd}$. As a consequence of the constraints on $\psi$, the 
equivalent function $\omega_{bd}$ has precisely the kind of behaviour 
needed to make $\omega_{bd} ({\rm today}) > 500$ in hyperextended 
inflation \cite{stein, quiros}. 

All of these features are generic and model independent, and arise as
consequences of the constraints on $\psi$. A highly 
non trivial aspect is that these constraints were originally derived 
in \cite{knobh} by the simple requirements that the 
graviton-dilaton theory be able to predict non trivial PPN parameters 
$\beta$ and $\gamma$ for a charge neutral point star, without any naked 
singularities.  

It is interesting to note that the big bang singularity is also avoided 
in the models of \cite{satya,rb} by totally different methods. In 
\cite{satya}, a Higgs potential for a complex BD type scalar results 
in a repulsive gravity which facilitates the absence of singularity. 
In \cite{rb}, a limiting curvature hypothesis is invoked to impose 
an upper bound (of Planckian magnitude) on the curvature which leads 
to the absence of singularity. 

This letter is organised as follows. Starting from the graviton-dilaton  
action, with an arbitrary function $\psi(\phi)$ describing the theory 
mentioned above, we present solutions \cite{myers,refr,refr2} 
of a homogeneous isotropic universe, and the expressions  
for various physical quantities. We then discuss the constraints on 
$\psi$, and their consequences on the physical quantities. Next, we add 
a dilatonic potential and repeat the above analysis. We then 
point out a surprising feature, which is relevent to the late stages of 
hyperextended inflation, and conclude with a few remarks. 

\vspace{4ex}

{\bf 2.} Following Dicke's approach \cite{dicke}, and as explained in 
detail in \cite{knobh}, the most general action for graviton and dilaton 
including the world line action for test particles, at least some of 
which have non zero rest mass, can be written as 
\begin{equation}\label{sein}
S = \int d^4 x \sqrt{- g} 
\left( R + \frac{1}{2} (\nabla \phi)^2 \right) 
+ \sum_i m_i \int \left( - e^{- \psi} g_{\mu \nu} 
d x^{\mu}_i d x^{\nu}_i \right)^{\frac{1}{2}} \; , 
\end{equation}
where $m_i$ are constants at least some of which are non zero, the sum is 
over different types of test particles and, with the dilaton potential 
set to zero, the arbitrary function $\psi (\phi)$ characterises our 
graviton-dilaton theory. $\psi$ cannot be gotten rid of, except when 
{\em all} test particles have zero rest mass which, by assumption, is 
not the case here. In our notation, the signature of the metric is 
$(- + + +)$ and $R_{\mu \nu \lambda \tau} = \frac{1}{2} \frac{\partial^2 
g_{\mu \lambda}} {\partial x^{\nu} \partial x^{\tau}} + \cdots$. 
Note that in (\ref{sein}) the curvature scalar $R$ appears canonically. 
For this reason, $g_{\mu \nu}$ is often referred to as `Einstein 
metric'. However, the test particles couple to dilaton now and feel both 
the gravitational and the dilatonic forces. Hence, they do not fall 
freely along the geodesics of $g_{\mu \nu}$. See \cite{dicke}. 

The action in (\ref{sein}) can be written equivalently in terms of 
the metric 
\begin{equation}\label{conf}
\hat{g}_{\mu \nu} = e^{- \psi} g_{\mu \nu} \; .
\end{equation}
It then becomes 
\begin{equation}\label{spsi}
S = \int d^4 x \sqrt{- \hat{g}} e^{\psi} 
\left( \hat{R} - \omega (\hat{\nabla} \phi)^2 \right) 
+ \sum_i m_i \int \left( - \hat{g}_{\mu \nu} 
d x^{\mu}_i d x^{\nu}_i \right)^{\frac{1}{2}} \; , 
\end{equation}
where $\omega = \frac{1}{2} (3 \psi_{(1)}^2 - 1)$. Here 
$\psi_{(n)} \equiv \frac{d^n \psi}{d \phi^n}$, the $n^{th}$ derivative 
of $\psi$ with respect to $\phi$. In (\ref{spsi}) and in the following, 
hats denote quantities involving $\hat{g}_{\mu \nu}$. Note that in 
(\ref{spsi}) the curvature scalar $\hat{R}$ does not appear canonically. 
However, the test particles now couple to the metric only canonically 
and, hence, fall freely along the geodesics of $\hat{g}_{\mu \nu}$. For 
this reason, we refer to $\hat{g}_{\mu \nu}$ as physical metric: since 
the test particles follow its geodesics, the quantities related to 
$\hat{g}_{\mu \nu}$ are the physically relevent ones. This is 
the original approach of Dicke \cite{dicke}. 

Thus, our theory is specified by the action given in (\ref{sein}) 
or, equivalently, in (\ref{spsi}) for graviton, dilaton, and for 
the test particles. It is characterised by one arbitrary function 
$\psi (\phi)$. Note that setting 
$\psi = \phi \; (3 + 2 \omega_{bd})^{- \frac{1}{2}}$ in (\ref{spsi}) 
one gets the Brans-Dicke theory; and, setting 
$\psi = \phi$ in (\ref{spsi}) one gets the graviton-dilaton part of 
the low energy string theory\footnote{modulo the choice of the test 
particle coupling. In the string theory literature, the action in 
(\ref{sein}), but with $\psi = 0$, and that in (\ref{spsi}), but with 
$\psi = \phi$, have both been used often.}. 

We now consider solutions describing a homogeneous isotropic 
universe. We first solve for $\phi$ and $g_{\mu \nu}$ using 
(\ref{sein}), and then obtain the physical metric $\hat{g}_{\mu \nu}$ 
using (\ref{conf}). The physical curvature scalar $\hat{R}$ is given by  
\begin{equation}\label{rhat}
\hat{R} = e^{\psi} (R + \frac{3}{2} (\nabla \psi)^2 
- 3 \nabla^2 \psi) \; . 
\end{equation}
The equations of motion obtained from 
(\ref{sein}) are 
\[
2 R_{\mu \nu} + \nabla_{\mu} \phi \nabla_{\nu} \phi = 
\nabla^2 \phi = 0 \; .
\]
With $\phi = \phi(t)$, the equations of motion in the gauge \\ 
$d s^2 = - d t^2 + a^2 (t) ( \frac{d r^2}{1 - k r^2} 
+ r^2 d \Omega^2 )$, where $k = 0, \; \pm 1$ and $d \Omega^2$ is 
the line element on an unit sphere, become 
\begin{equation}\label{eqn}
\frac{6 \ddot{a}}{a} + \dot{\phi}^2 =  
\frac{2 \ddot{a}}{a} + \frac{4 (\dot{a}^2 + k)}{a^2} = 
\ddot{\phi} + \frac{3 \dot{a} \dot{\phi}}{a} = 0 \; , 
\end{equation}
where upper dots denote $t$-derivatives. Following the original work 
of \cite{myers} on string theoretic cosmology, a lot of work has been 
carried out in this arena in which equations (\ref{eqn}), and their 
higher dimensional and other generalisations, have been solved 
\cite{refr,refr2}. The solutions to (\ref{eqn}) are of the form 
\begin{equation}\label{soln}
a = A t^n \; , \; \; 
e^{\phi - \phi_0} = t^{\epsilon m} \; , 
\end{equation}
where $A$ and $\phi_0$ are constants, $\epsilon = \pm 1$ and, 
by definition, $m$ is positive. We will set $\phi_0 = 0$ without any 
loss of generality. Equation (\ref{eqn}) gives
\begin{equation}\label{nm0}
(n, m) = (\frac{1}{3}, \frac{2}{\sqrt{3}}) \; \; , \; \; k = 0 \; , 
\end{equation}
where the square root, above and in the following, is to be taken with 
a positive sign. The physical curvature scalar $\hat{R}$ is given by 
\begin{equation}\label{rhat1}
\hat{R} = (6 \psi_{(2)} - 3 \psi_{(1)}^2 + 1) e^{\psi} R 
\; ; \; \; \; \; R = \frac{6 n (1 - 2 n)}{t^2} - \frac{6 k}{a^2} \; . 
\end{equation}
The parameter $t$ will vary only from $0$ to $\infty$, since the dilaton 
$\phi$ becomes complex for $t < 0$ whose interpretation is not clear. 

The physical metric $\hat{g}_{\mu \nu}$, using (\ref{conf}), becomes 
\begin{equation}\label{metric}
d \hat{s}^2 = - d \hat{t}^2 + \hat{a}^2 
\left( \frac{d r^2}{1 - k r^2} + r^2 d \Omega^2 \right) \; , 
\end{equation}
where $\hat{t}$ is the physical time and $\hat{a}$ is the physical 
scale factor, given by 
\begin{equation}\label{hats}
\frac{d \hat{t}}{d t} = e^{- \frac{\psi}{2}} \; , \; \; 
\hat{a} = e^{- \frac{\psi}{2}} a \; . 
\end{equation}
In low energy string theory, $\psi = \phi$. Using the above solutions, 
we get 
\[
\hat{t} = \frac{\sqrt{3}}{\sqrt{3} - \epsilon} \; 
t^{1 - \frac{\epsilon}{\sqrt{3}}} + constant \; , \; \;  
\hat{R} = - \frac{4}{3} \; t^{- 2 (1 - \frac{\epsilon}{\sqrt{3}})} \; . 
\]
It follows that as $t \to 0$, the physical time $\hat{t} \to constant$, 
and the physical curvature scalar $\hat{R} \to \infty$. Thus there is 
a physical curvature singularity at a finite time in the past, beyond 
which the physical time $\hat{t}$ cannot be continued. Similarly, it can 
be seen that this singularity, referred to as big bang singularity in the 
following, is also present in BD theory. 

\vspace{4ex}

{\bf 3.} There is another aspect of BD and low energy string theories, 
shown in \cite{kppn}: they cannot 
predict non trivial values of the PPN parameters $\beta$ 
and/or $\gamma$ for a charge neutral point star, without any naked
singularities. In \cite{knobh} we looked for, and obtained, 
a graviton-dilaton theory of the form given in (\ref{spsi}), 
requiring it only to be able to 
predict non trivial values for $\beta$ and/or $\gamma$  for 
a charge neutral point star, without any naked singularities. These 
simple requirements imposed certain constraints on $\psi$, to be 
described presently. For the static spherically symmetric case, these 
constraints led to the novel features of the gravitational force 
becoming repulsive at distances of the order of, but greater than, 
the Schwarzschild radius $r_0$ and the absence of a horizon for 
$r > r_0$, suggesting that black holes are 
unlikely to form in a stellar collapse in this theory. 

We will now consider this theory, which is defined by any function 
$\psi(\phi)$ satisfying the following constraints (see \cite{knobh} for 
details) :   
\begin{eqnarray}
{\rm (i)} & & 
\bar{a} - \bar{b} \psi_{(1)} (\bar{\phi}) > 0 \; , \; \; 
\bar{b}^2 \psi_{(2)} (\bar{\phi}) = \delta \; , \; \; 
\bar{b} \psi_{(1)} (\bar{\phi}) = \epsilon \; , \label{c1} \\
{\rm (ii)} & & 
\psi_{(n)} (\phi) \equiv \frac{d^n \psi}{d \phi^n} 
= (finite) \; \; \; \; \forall \; n \ge 1 
\; , \; \; - \infty \le \phi \le \infty \; , \label{c2} \\ 
{\rm (iii)} & & 
\lim_{\phi \to \pm \infty} \psi = -  \lambda |\phi| 
\; \; , \; \; \; \lambda \ge \sqrt{3} \; . \label{c3}
\end{eqnarray}
In equation (\ref{c1}), $\bar{\phi}$ 
is the asymptotic value of $\phi$ which can be set to zero, $\delta$ 
and $\epsilon$ are $< 10^{- 3}$ with at least one of them nonzero, 
$0 < \bar{a} < 1$, and $\bar{b} = \pm \sqrt{1 - \bar{a}^2} \ne 0$. 
Note that equations (\ref{c2}) and (\ref{c3}) imply a finite upper 
bound on $\psi$, {\em i.e.} $\psi \le \psi_{max} < \infty$. 

There are many functions $\psi$ satisfying these requirements, 
{\em e.g.}\ $\psi = - \lambda \sqrt{(\phi - \phi_1)^2 + c^2}$ 
where $\phi_1$ and $c^2$ are 
constants. However, an explicit form of $\psi$ is not needed, since 
we are concerned here only with generic, model independent features, 
true for any $\psi$ satisfying only the constraints (i)-(iii). 
We will now consider these features. 

From equations (\ref{soln}) and (\ref{c3}), it follows that 
\begin{equation}\label{psit}
e^{\psi} \to t^{- \lambda m} \; \; {\rm as} \; \; 
t \to \infty \; , \; \; {\rm and} \; \; 
e^{\psi} \to t^{\lambda m} \; \; {\rm as} \; \; t \to 0 \; .
\end{equation}
Note that, by definition, $m$ is positive. Also, 
$\psi \le \psi_{max} < \infty$. Therefore, 
$e^{- \frac{\psi}{2}} \ge e^{- \frac{\psi_{max}}{2}} > 0$. 
It then follows from (\ref{hats}) that 
the physical time $\hat{t}$ is a strictly increasing function of $t$, 
given by 
\begin{equation}\label{tphys}
\hat{t} - \hat{t}_0 = \int_{t_0}^t d t \; e^{- \frac{\psi}{2}} \; , 
\end{equation}
where $t_0$ and $\hat{t}_0$ are some finite 
constants, which are not relevent 
for our purposes. Using equations (\ref{rhat1}) and (\ref{hats}) we 
obtain the following result, where $\hat{t}$ is defined upto a finite 
additive constant. In the limit $t \to \infty$, 
\begin{equation}\label{t1}
\hat{t} = \frac{2}{2 + \lambda m} t^{1 + \frac{\lambda m}{2}} 
\; , \; \; 
\hat{a} = A t^{n + \frac{\lambda m}{2}} \; , \; \; 
\hat{R} = \frac{2}{3} \; (1 - 3 \lambda^2) \; 
t^{- (2 + \lambda m)} \; . 
\end{equation}
And, in the limit $t \to 0$, 
\begin{equation}\label{t0}
\hat{t} = \frac{2}{2 - \lambda m} t^{1 - \frac{\lambda m}{2}} 
\; , \; \; 
\hat{a} = A t^{n - \frac{\lambda m}{2}} \; , \; \; 
\hat{R} = \frac{2}{3} \; (1 - 3 \lambda^2) \; 
t^{- (2 - \lambda m)} \; . 
\end{equation}
In equation (\ref{t0}), $\hat{t} = \ln t$ if $\lambda m = 2$. 

The behaviour given in equations (\ref{psit}), (\ref{t1}), and 
(\ref{t0}) is generic and model independent, true for any function 
$\psi$ satisfying the constraints (i)-(iii). It follows from 
(\ref{t1}) that as $t \to \infty$, 
\begin{equation}\label{tinf}
\hat{t} \to \infty \; , \; \; 
\hat{a} \to \infty \; , \; \; 
\hat{R} \to 0 \; , 
\end{equation}
independently of the values of $n, \; m$, and $\lambda$, which are all 
positive. 

However, the behaviour of these physical quantities as $t \to 0$, where  
there is a potential big bang singularity, will depend on the values of 
$n, \; m$, and $\lambda$. Using the solution (\ref{nm0}) and the 
constraint $\lambda \ge \sqrt{3}$, we have $\lambda m \ge 2$. 
As a beautiful consequence of this inequality,  
it follows from (\ref{t0}) that, as $t \to 0$, 
\begin{equation}\label{tzero}
\hat{t} \to - \infty \; , \; \; 
\hat{a} \to \infty \; , \; \; 
\hat{R} \to 0 \; \; (or, \; constant) \; , 
\end{equation}
Hence, as $t \to 0$, the physical time $\hat{t}$ can be continued 
indefinitely into the past, the physical scale factor of the universe 
becomes infinite, and the physical curvature scalar becomes zero (or, 
approaches a constant). Also, since $\psi \le \psi_{max} < \infty$, 
it is clear from equation (\ref{hats}) that the physical scale factor 
$\hat{a}$ never vanishes at any time $\hat{t}$ and, from equation 
(\ref{rhat1}), that the physical curvature scalar remains finite for 
all $\hat{t}$, suggesting the absence of big bang singularity. 

However, for the singularity to be absent, {\em all} other curvature   
invariants must also be finite. As shown in the appendix, any curvature 
invariant is of the form given in (\ref{app1}). Since equation  
(\ref{c2}) is satisfied, and since $(e^{\psi} t^{- 2})$ is finite 
for all $\hat{t}$ as can be seen from the above discussions, 
it follows that {\em all} the $n^{th}$ order curvature invariants 
are also finite, for all $\hat{t}$. This shows that the big bang 
singularity, shown earlier to be present in BD and low energy string 
theories, is indeed absent here. 

Notice that these features are generic and model independent, true for 
any function $\psi$ satisfying only the constraints (i)-(iii). Also note 
the crucial role of the constraint $\lambda \ge \sqrt{3}$, which 
beautifully ensures that the physical time $\hat{t}$ can be continued 
indefinitely into the past and that all the curvature invariants are 
finite for all $\hat{t}$. This is a highly non trivial aspect of the 
graviton-dilaton theory considered here since the constraints (i)-(iii), 
which ensure the absence of big bang singularity here, originate from 
a totally different  requirement - that the graviton-dilaton theory be 
able to predict non trivial PPN parameters $\beta$ and $\gamma$, 
obtained from static spherically symmetric solutions, for a charge 
neutral point star without any naked singularities.  

\vspace{4ex}

{\bf 4.} As another illustration of the generic cosmological features 
of the graviton-dilaton theory presented here, 
consider the action (\ref{spsi}) with a dilaton potential, 
$V(\phi) = \Lambda e^{\psi + \alpha \phi}$, coupled as follows:  
\begin{equation}\label{sc}
S = \int d^4 x \sqrt{- \hat{g}} e^{\psi} \left( \hat{R} 
- \omega (\hat{\nabla} \phi)^2  + \Lambda e^{\psi + \alpha \phi} 
\right) \; . 
\end{equation}
where $\Lambda$ is a positive constant,  
and $0 < \alpha^2 \le 3$. If $\alpha = 0$ 
then in the solutions, $\phi$ and, hence, $\psi$ are constants and the 
non trivial consequences  of our theory cannot be seen. Hence, we take 
$\alpha \ne 0$. The upper bound on $\alpha^2$ is set so that the 
potential $V \equiv \Lambda e^{\psi + \alpha \phi}$ is finite for 
all $\phi$, which is ensured by the constraint (iii). 

Proceeding as before to the Einstein frame using (\ref{conf}), the 
equations of motion for a homogeneous isotropic universe are 
\begin{equation}\label{ceqn}
\frac{6 \ddot{a}}{a} + \dot{\phi}^2 =  
\frac{2 \ddot{a}}{a} + \frac{4 (\dot{a}^2 + k)}{a^2} = 
- \frac{1}{\alpha} \left( \ddot{\phi} 
+ \frac{3 \dot{a} \dot{\phi}}{a} \right) = \Lambda e^{\alpha \phi} \; .  
\end{equation}
Such equations arising from dilaton potential and their solutions have 
also been considered in \cite{myers,refr2} for specific values of 
$\alpha$. 
The solutions to (\ref{ceqn}) give $a$ and $\phi$ of the same form as 
in (\ref{soln}), with 
\begin{equation}\label{cnm0}
(n, \epsilon m) = (\frac{1}{\alpha^2}, - \frac{2}{\alpha}) \; \; , \; \;  
\Lambda e^{\alpha \phi_0} = \frac{2 (3 - \alpha^2)}{\alpha^4} 
\; \; , \; \; k = 0 \; , 
\end{equation} 
or  
\begin{equation}\label{cnm1}
(n, \epsilon m) = (1, - \frac{2}{\alpha}) \; \; , \; \;  
\Lambda e^{\alpha \phi_0} = \frac{4}{\alpha^4} \; \; , \; \; 
A^2 =  \frac{k \alpha^2}{1 - \alpha^2} \; , 
\end{equation} 
where $\epsilon = - sign(\alpha)$ so that $m \equiv \frac{2}{|\alpha|}$  
is positive and, in equation (\ref{cnm1}), $k = sign(1 - \alpha^2)$. 
If $\alpha^2 = 1$ then $k = 0$ and the constant $A$ is arbitrary, and 
equations (\ref{cnm0}) apply. The physical 
curvature scalar $\hat{R}$ is given by equation (\ref{rhat1}). 
Analysing, as before, the physical quantities in these solutions we 
find, for a generic choice of $\alpha$ with $0 < \alpha^2 \le 3$, that 
there are curvature singularities in BD and low energy string theories, 
where the physical curvature scalar $\hat{R} \to \infty$ at some 
physical time $\hat{t}$. 

Now consider 
the theory given by action (\ref{sc}), with the function $\psi$ 
obeying the constraints (i)-(iii). The physical quantities $\hat{t}, \;  
\hat{a}$, and $\hat{R}$ can be analysed as before. Equations 
(\ref{psit})-(\ref{t0}) apply, leading, as $t \to \infty$, to the 
behaviour given in (\ref{tinf}). To find the behaviour as $t \to 0$,   
the values of $(n, m)$ given in equations (\ref{cnm0}) and (\ref{cnm1}) 
are to be used. Since $m = \frac{2}{|\alpha|}, \; \alpha^2 \le 3$ and 
$\lambda \ge \sqrt{3}$, we have again $\lambda m \ge 2$. 
Therefore, as $t \to 0$, 
\begin{equation}\label{ctzero}
\hat{t} \to - \infty \; , \; \; 
\hat{R} \to 0 \; \; (or, constant) \; , 
\end{equation}
and the physical scale factor $\hat{a}$ will diverge if $n < 3$, 
or equivalently if $\alpha^2 > \frac{1}{3}$, and vanish if $n > 3$, 
or equivalently if $\alpha^2 < \frac{1}{3}$. 
Very importantly, however, the physical time $\hat{t}$ can be continued   
indefinitely into the past or future, and the physical curvature scalar 
$\hat{R}$ and, hence, following the same analysis as before, 
all other physical curvature invariants also 
are finite for all $\hat{t}$. This again shows that there is no big bang  
singularity in the theory given by the action (\ref{sc}), 
which has a non trivial dilaton potential. 

\vspace{4ex}

{\bf 5.} Another novel feature of the graviton-dilaton action 
(\ref{spsi}) follows by rewriting it as   
\begin{equation}\label{heinf}
S = \int d^4 x \sqrt{- \hat{g}} e^{\psi} \left( \hat{R} 
+ \omega_{bd} (\hat{\nabla} \psi)^2  \right) \; , 
\end{equation}
where $\omega_{bd} = \frac{1 - 3 \psi_{1}^2}{2 \psi_{1}^2}$ 
is the BD function 
which, if constant, is the standard BD parameter\footnote{
In $\omega_{bd}$, the function $\psi_{1}^2$ is to be expressed only 
in terms of $\psi$. For example, 
if $\psi = - \lambda \sqrt{(\phi - \phi_1)^2 + c^2}$ then 
$\psi^2_{1} = \frac{\lambda^2 (\psi^2 - \lambda^2 c^2)}{\psi^2}$.}. 
Also, the standard BD scalar field is related to $\psi$ 
by $\Phi_{bd} \equiv e^{\psi}$. 

As noted before, the constraints (i)-(iii) on $\psi$ imply a finite 
upper bound $\psi_{max}$ on $\psi$. Hence, $\psi_{(1)} = 0$ for at 
least one finite value of $\phi \equiv \phi_c$, where 
$\psi(\phi_c) = \psi_{max}$. Therefore, 
as $\phi \to \phi_c, \; \psi_{(1)} \to 0$, and 
$\omega_{bd} \to  \frac{1}{2 \psi_{1}^2}$, 
which is precisely the kind of behaviour needed to make 
$\omega_{bd} ({\rm today}) > 500$ in hyperextended inflation 
\cite{stein, quiros}. The example of 
$\psi = - \lambda \sqrt{(\phi - \phi_1)^2 + c^2}$ 
corresponds, as $\phi \to \phi_c \; ( = \phi_1 )$, 
to the model of \cite{quiros} with $\alpha = 1$; 
the examples of $\psi$ corresponding, as $\phi \to \phi_c$, 
to other values of $\alpha$ in \cite{quiros} can be easily found. 

This novel feature, that $\omega_{bd}$ here has precisely the kind of 
behaviour needed to make $\omega_{bd} ({\rm today}) > 500$ in 
hyperextended inflation, is a generic and model 
independent consequence of the constraints (i)-(iii). We find it quite 
surprising that this feature emerges naturally in the graviton-dilaton 
theory considered here. The question of whether other aspects of 
hyperextended inflation also follow in this theory is presently under 
study \cite{ks}.  

\vspace{4ex}

{\bf 6.} We comment now on some aspects of the present theory which 
can be studied further. Note that we have not included here matter and 
radiation (except as test particles which do not affect the fields), 
which is necessary for cosmology. What is the effect of their 
inclusion? Our preliminary studies on this question 
indicate that, even with matter and/or radiation present, the big 
bang singularity may still be absent. Work on this is in progress. 

Does the present theory stem from a fundamental theory, {\em e.g.}\ 
string theory? We do not know the answer. The only connection, if 
at all, to a fundamental theory that we can see is the following. 
$4 + 1$-dimensional Kaluza-Klein gravity compactified on a circle to 
$3 + 1$ dimensions gives a graviton-dilaton theory with 
$\psi = \sqrt{3} \phi$ (see \cite{gm} for example). Equation (\ref{c3}) 
is then barely satisfied when $\phi \to - \infty$, but not when 
$\phi \to \infty$. Now if there is some connection between four and five 
dimensional theories in some limit (say, as $\phi \to - \infty$),  
analogous to the one recently discovered 
between ten and eleven dimensional string theories \cite{witten}, then 
there is just a possibility, admittedly far fetched, that one may be able 
to derive the present theory from a fundamental one.  

We will end this letter with one irresistible speculation. 
The short distance repulsion that seems to be present in this theory 
may perhaps soften, and even eliminate, the ultraviolet divergences in 
quantum gravity. However, even if this were the case, the repulsive 
effects come into play only at short distances where the fields are 
often strong and where the known perturbative techniques may break down. 
Hence, a study of this important issue is likely to be difficult. 

Thus, keeping in view various novel features of the present theory seen 
in \cite{knobh} and in this letter, we believe its further study to be 
fruitful. 

\vspace{2ex} 

{\bf Acknowledgements:} 

It is a pleasure to thank S. Ghosh, H. S. Mani, and T. R. Seshadri for 
many helpful discussions. We are particulary greatful to the referee 
for pointing out a major flaw in our original proof of the absence of 
singularities - correcting this flaw led to the more stringent 
constraint on $\psi$ given in (\ref{c2}), for informing us of 
the references \cite{refr,refr2}, and for pointing out some 
mistakes in the original version. 

\vspace{4ex}

\begin{center}
{\bf Appendix}
\end{center}

All curvature invariants can be constructed using metric tensor, 
Riemann tensor, and covariant derivatives which contain ordinary 
derivatives and Christoffel symbols $\hat{\Gamma}_{a b}^c$. When 
the metric is diagonal, every term in any curvature invariant can be 
grouped into factors, each of which is of one of the following forms 
(no summation over repeated indices): (A) $\sqrt{\hat{g}^{a a} 
\hat{g}^{b b} \hat{g}^{c c} \hat{g}^{d d}} \hat{R}_{a b c d}$, 
(B) $\sqrt{\hat{g}^{a a} \hat{g}^{b b} \hat{g}_{c c}} 
\hat{\Gamma}_{a b}^c$, or (C) $\sqrt{\hat{g}^{a a}} \partial_a$. 

Taking $\hat{g}_{\mu \nu}$ given in (\ref{metric}), and the solutions 
for $a$, and $\phi$ given in (\ref{soln}), the above forms can be 
calculated explicitly. The calculation is straightforward, and 
the result is that (A), (B), and (C) can be written, symbolically, as 
\begin{eqnarray*}
& & (A) \simeq U e^{\psi} t^{- 2} \; , \; \; \; 
(B) \simeq V ( e^{\psi} t^{- 2} )^{\frac{1}{2}} \; ,  \\
& & (C)^n \cdot (A)^p (B)^q \simeq W_{n + 2} 
( e^{\psi} t^{- 2} )^{p + \frac{1}{2}(q + n)} \; , 
\end{eqnarray*}
where $U$ and $V$ are functions of $t, \; \psi_{(1)}$, and $\psi_{(2)}$ 
only, and $W_{k}$ are functions of $t$ and 
$\psi_{(l)}, \; \; 1 \le l \le k$. It turns out, as a result of 
the way various factors are grouped, that the explicit $t$-dependent 
parts in $U, \; V$, and $W_k$'s are finite at $t = 0$ (in fact, they 
have divergence at $t = \infty$ only which is mild and such that it is 
suppressed by the accompanying $(e^{\psi} t^{- 2})$ factors). 

Hence, any curvature invariant constructed from $m$ Riemann tensors, 
$n$ covariant derivatives, and the requisite number of metric tensors 
will be of the form 
\begin{equation}\label{app1}
\tilde{W} (t; \psi_{(1)}, \psi_{(2)}, \cdots, \psi_{(n + 2)}) \; 
( e^{\psi} t^{- 2} )^{m + \frac{n}{2}} \; 
\end{equation}
where the explicit $t$-dependent parts of $\tilde{W}$ are finite at 
$t = 0$ (in fact, they have divergence at $t = \infty$ only which is 
mild and such that it is suppressed by the accompanying 
$(e^{\psi} t^{- 2})$ factors). Note, as 
an example, that the curvature scalar given in (\ref{rhat1}) belongs 
to type (A), and has the above form with $m = 1$ and $n = 0$. 

\vspace{2ex} 

{\bf Note Added in Proof:} 

There is an issue which we are unable to resolve as yet: whether or not 
a test photon, or a massless test particle, sees a singularity in 
the graviton-dilaton background described here. 

On the one hand, the physical metric $\hat{g}_{\mu \nu}$ can be 
conformally transformed to $g_{\mu \nu}$, which is singular - suggesting 
that a test photon sees a singularity. On the other hand, however, 
(i) the conformal factor is singular, so the validity of 
this transformation is not automatic; (ii) our calculations, the details 
of which will be given in \cite{ks}, show that the big bang singularity 
is absent and the space time described by $\hat{g}_{\mu \nu}$ is 
regular everywhere even upon including macroscopic amount of matter 
and/or radiation, that modify the space time. This would suggest that 
a test photon should see no singularity. (iii) Moreover, in $3 + 1$ 
dimensional space time and within the context of Dicke type ansatz, 
the natural photon-dilaton coupling is ambiguous, and appears to call 
for another arbitrary function of dilaton. If so, this function may be 
quite relevent to the present issue. 

For these reasons, we are unable to resolve the above issue definitively. 


\end{document}